# Does the repulsive interatomic potential determine fragility in metallic liquids?

Christopher E. Pueblo[1,2], Minhua Sun[2,3] and Kenneth F. Kelton[*,1,2]

[1]*Department of Physics, Washington University, St. Louis, MO 63130, USA.* [2]*Institute of Materials Science and Engineering, Washington University, St. Louis, MO 63130, USA.*

[3]*Department of Physics, Harbin Normal University, Harbin, Heilongjiang Province 150025, China.*

**The dynamical behavior of liquids is frequently characterized by the fragility, which can be defined from the temperature dependence of the shear viscosity, $\eta$. For a strong liquid, the activation energy for $\eta$ changes little with cooling towards the glass transition temperature, $T_g$. The change is much greater in fragile liquids, with the activation energy becoming very large near $T_g$. While fragility is widely recognized as an important concept, for example believed to play an important role in glass formation, the microscopic origin of fragility is poorly understood. Here, we present new experimental evidence showing that fragility reflects the strength of the repulsive part of the interatomic potential, which can be determined from the steepness of the pair distribution function near the hard-sphere cutoff. Based on an analysis of scattering data from ten different metallic alloy liquids, we show that stronger liquids have steeper repulsive potentials.**

Almost one-half century ago, Polk and Turnbull argued for a connection between the interatomic potential and dynamics, proposing that the viscosity should be larger in liquids composed of "hard" atoms (with a steep repulsive potential) than in liquids containing "softer" atoms.[1] Krausser *et al.*[2] recently recast this argument into a correlation between the repulsive potential and fragility. By relating the temperature dependence of the infinite-frequency shear modulus, $G_{inf}$, in a few metallic glasses to the steepness, $\lambda$ of the repulsive part of the interatomic potential, they obtained results that are in disagreement with expectations from Polk and Turnbull's arguments. However, their results are in agreement with some theoretical studies[3] and with some studies of colloidal suspensions, where the relaxation times as a function of concentration in a suspension of "stiff" particles show fragile behavior, while suspensions of "soft" particles are strong.[4] There are prominent points of disagreement with this interpretation, however. For example, Casalini *et al*. argue that fragilities calculated as a function of concentration in colloidal suspensions are not strictly analogous to those calculated from relaxation times as a function of temperature and pressure.[5] This is supported by the results from their experimental studies, which indicate that strong liquids have steeper interaction potentials,[6] in agreement with other theoretical studies.[7-10] In this letter, we present new data in metallic liquids that agree with the expectations of Polk and Turnbull and with Casalini's results.

A starting point of Krausser *et al*.'s development of an expression for $G_{inf}$ is that the steepness of the repulsive potential is reflected in the slope of the low-r side (near the cutoff radius) of the pair distribution function, $g(r)$.[2] Here, we present new direct experimental measurements of these slopes for ten metallic liquids (some, but not all, glass forming) and correlate them with the fragility, obtained from the measured viscosity. The results demonstrate that stronger liquids have steeper repulsive potentials, and that the steepness is a function of the temperature of the liquid. As will be shown, both of these experimental observations are in agreement with results from a molecular dynamics simulation of a representative Cu-Zr liquid.

The potential of mean force between two atoms, $U_m$, (which is typically used as a first-approximation to the interaction potential,[11]) is defined in terms of $g(r)$[12]

$$\frac{U_m}{k_B T} = -\ln(g(r)), \tag{1}$$

where $k_B$ is Boltzmann's Constant and $T$ is the temperature. The low-r limit of g(r) can be expressed as[2]

$$g(r) = C(r - \sigma)^\lambda, \quad (2)$$

where $C$ is a fitting constant, $\sigma$ is the average ionic core diameter (calculated from a weighted average of the liquid's constituent elemental ionic core diameters collected from the literature[13]), and $\lambda$ is the steepness of the effective interaction potential. The pair distribution function is obtained from the Fourier transform of the structure factor, $S(q)$, which is calculated from X-ray scattering results data on containerlessly-processed metallic liquids (see "Methods" for a discussion of the experimental procedure and the method of data reduction). For illustration, fig. 1.a shows two $g(r)$'s, collected from the same liquid alloy at different temperatures. The low-$r$ portions of the peaks are fit to eq. 2 (a magnified portion of the low-r region is shown in the inset to the figure to illustrate the quality of the fits to the data). As shown in fig. 1.b, $\lambda$ is clearly temperature-dependent, which is in opposition with previous arguments.[14] The increasing value of $\lambda$ with decreasing temperature reflects the sharpening of the first peak of $g(r)$ with structural ordering.

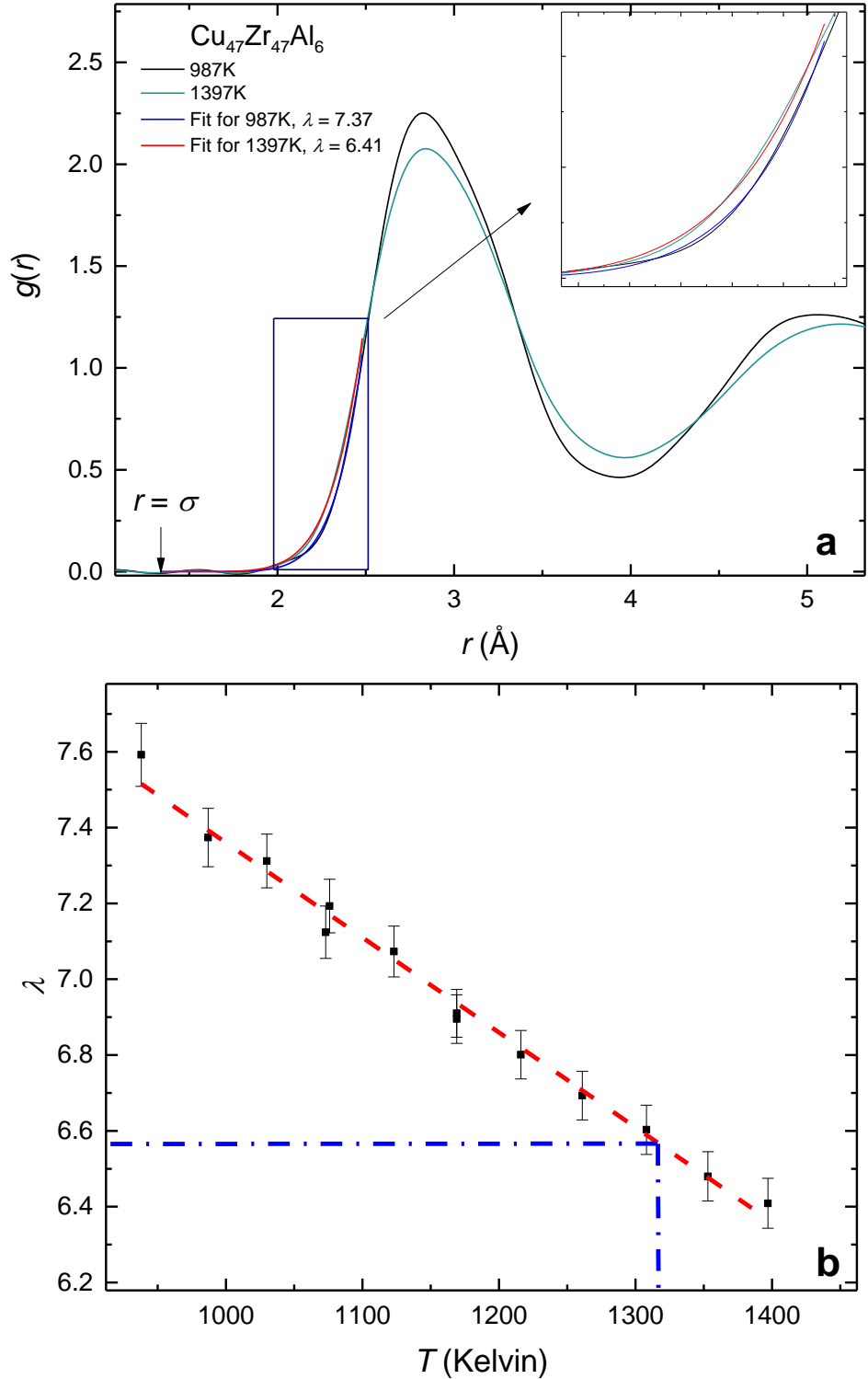

**Figure 1** – (a) Pair distribution function, $g(r)$, obtained from X-ray scattering experiments for both the equilibrium (1397K) and the supercooled (987K) $Cu_{47}Zr_{47}Al_6$ liquid ($T_l$ = 1172K), showing the fit of the low-r side of the first peak to eq. 2. The top right inset shows an expanded view of the fit region. (b) The steepness $\lambda$ as a function of temperature for the $Cu_{47}Zr_{47}Al_6$ liquid. The blue line shows the temperature ($T_A$) at which $\lambda$ is evaluated for the liquids studied here.

Because $\lambda$ is temperature-dependent, a comparison between different liquids of different fragilities requires that it be evaluated at a common reference temperature. The glass transition temperature would be a suitable reference; however, this is not known for all of the liquids studied (also some of which cannot be made into a glass). Recent MD and experimental studies have identified a suitable high-temperature reference, $T_A$, the temperature at which the atomic dynamics underlying the shear viscosity, $\eta$, become cooperative[15] and where dominant local clusters in the liquid begin to connect.[16] This temperature has also been related to an avoided critical point, where geometric frustration prevents the liquid from undergoing a phase transition to the liquid's energetically preferred structure.[17] $T_A$ can be identified experimentally as the temperature where $\eta$ changes from Arrhenius to super-Arrhenius behavior. Studies suggest that the processes that result in the glass transition actually start with the onset of cooperativity at $T_A$. For metallic liquids the first experimental studies showed that an estimate of $T_g$ can be obtained from $T_A$, with $T_A \sim 2T_g$[20]; later studies have shown that while still close to two, the multiplicative factor for $T_g$ varies a little with fragility.[16, 18, 19 20] Fig. 2.a. shows measured viscosity data for a $Cu_{47}Zr_{47}Al_6$ liquid and the value obtained for $T_A$ (see "Methods" for a discussion of the viscosity measurement and the determination of $T_A$). Fragility is typically defined from an Angell plot ($\log_{10}(\eta)$ vs. $T_g/T$). However, since liquids that do not form glasses are also included in this study, it is not possible to obtain their values of $T_g$. Since $T_A \sim 2T_g$, a plot of $\log_{10}(\eta)$ vs. $T_A/T$ contains the same information as a plot of $\log_{10}(\eta)$ vs. $T_g/T$. For the evaluation of $\lambda$ and also for scaling the viscosity data to determine kinetic fragility in the liquids studied, $T_A$ is, therefore, used in lieu of $T_g$.

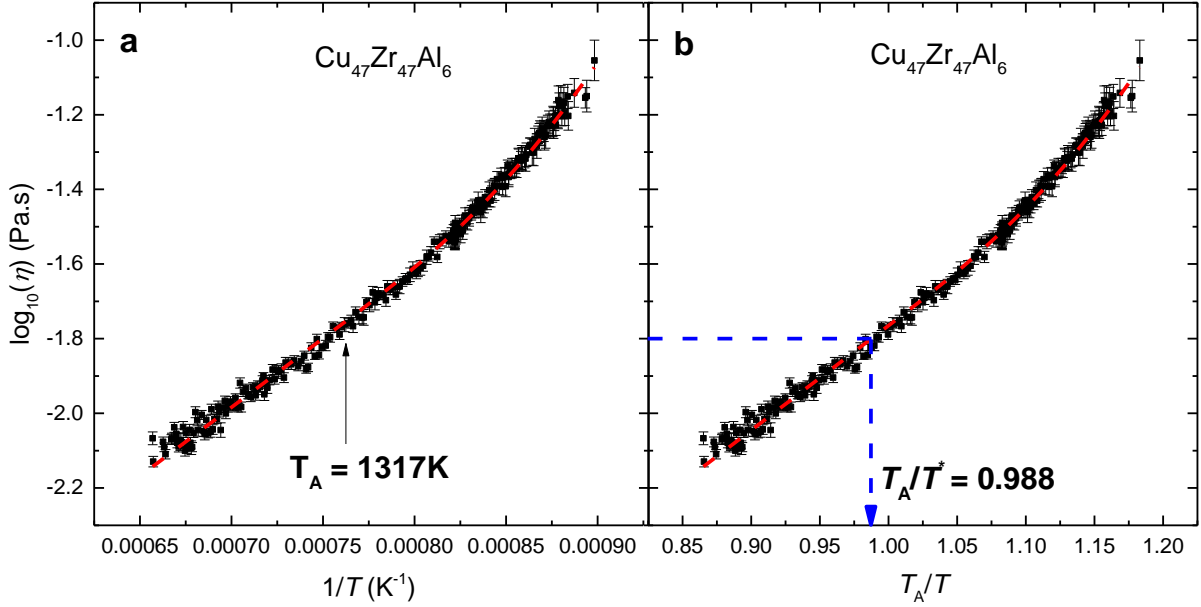

**Figure 2 -** (a) Viscosity for the equilibrium and supercooled $Cu_{47}Zr_{47}Al_6$ liquid, with fits shown (see methods for the expression used for the fits). (b) A modified fragility plot that scales the inverse temperature to $T_A$ instead of $T_g$. The kinetic fragility is determined by the reduced temperature $T_A/T^*$, where $T^*$ is the temperature at which $\log_{10}(\eta) = -1.8$.

Fragility is most commonly defined in terms of the fragility parameter, $m$,

$$m = \left.\frac{d\log_{10}\eta}{d(T_g/T)}\right|_{T=T_g}, \qquad (3)$$

with stronger liquids having a smaller $m$ (lower effective activation energy) near $T_g$ than fragile liquids. Since viscosity data for many of these liquids are not available near $T_g$, however, and because some cannot be made into glasses, another approach must be followed to compute the fragility. Angell and co-workers[21, 22] pointed out that since the magnitude of the viscosity at high temperature, relative to $T_g$, is larger for stronger liquids, the scaled temperature at which $\log_{10}(\eta)$ takes on a common value ($T_g/T^*$) can be used as a measure of fragility. They chose $T^*$ to be the temperature for which the common value for $\eta$ was midway between that at $T_g$ and the extrapolated value at infinite temperature, $\eta_o$. Other choices for the common value for $\eta$, however, are equally valid. We identify $T^*$ as the temperature where $\log_{10}(\eta) = -1.8$, which is experimentally accessible for all liquids studied. With this choice $T_A/T^*$ provides the measure of the liquid's fragility (fig.

2b); stronger liquids have smaller values for the ratio of $T_A/T^*$ than fragile liquids (as shown in fig. S1 in the supplemental information, in those glass forming liquids where $m$ and $T_g$ are available, $m$ correlates with both $T_g/T^*$ and $T_A/T^*$, indicating that either can be used as a fragility metric for high temperature liquids).

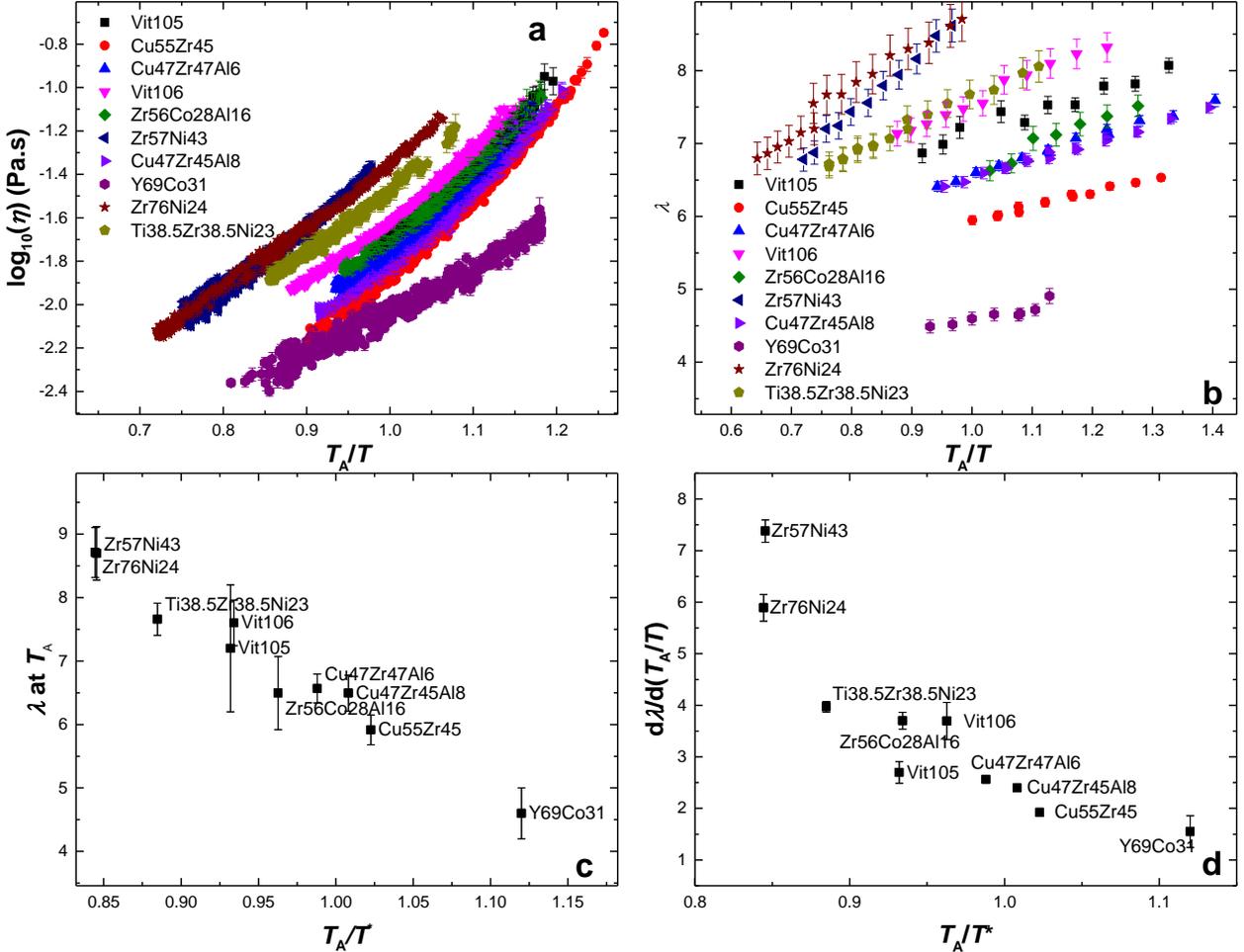

**Figure 3 -** (a) A modified Angell plot, showing the viscosity as a function of scaled inverse temperature, $T_A/T$, for all ten of the metallic liquids studied. They range from good glass formers (Vit106, $Zr_{56}Co_{28}Al_{16}$) to marginal glass formers ($Zr_{57}Ni_{43}$, $Zr_{76}Ni_{24}$, $Y_{69}Co_{31}$), and include one that does not form a glass ($Ti_{38.5}Zr_{38.5}Ni_{23}$). This shows that scaling to $T_A$, instead of $T_g$, allows a more diverse group of liquids to be characterized, rather than just those that easily vitrify. (b) A plot of $\lambda$, calculated from X-ray scattering data, versus $T_A/T$. The similarity between Figs. 3.a and 3.b suggests a connection between fragility and the interaction potential. (c) The value of $\lambda$ evaluated at $T_A$ as a function of the fragility parameter $T_A/T^*$. A correlation between the two quantities is clear, with stronger liquids having larger values of $\lambda$. (d) The derivative $d\lambda / d(T_A/T)$ as a function of the measure of fragility, $T_A/T^*$.

The logarithms of the viscosities of the ten metallic liquids studied are shown as a function of scaled inverse temperature, $T_A/T$ (fig. 3.a). The magnitude and temperature dependence of $\lambda$ mimic the magnitude and temperature dependence of the viscosity (fig. 3.b), suggesting a correlation between $\lambda$ and the kinetic fragility. This is confirmed in fig. 3.c, showing the values of $\lambda$ evaluated at $T_A$ as a function of $T_A/T^*$. These data demonstrate that stronger liquids have steeper effective repulsive potentials. The conclusion is the same if $T_g$ is used instead of $T_A$ for liquids that easily form glasses and for which $T_g$ can be accurately determined (see fig. S2 in the supplemental information). As shown in fig. 3.d, the rate of change of $\lambda$ in an Angell-type plot ($d\lambda/d(T_A/T)$) also tracks with the fragility, indicating that the nearest neighbor correlation is ordering more rapidly at high temperatures in stronger liquids.

As illustrated in the inset in fig. 1, the temperature dependence of $\lambda$ is constant over the range measured, while below $T_A$, the viscosity becomes super-Arrhenius. While this might appear incorrect, it is consistent with the functional form for the viscosity derived by Krausser et al. [2]

$$log_{10}(\eta/\eta_0) = \frac{C}{T} * \exp[(2+\lambda)\alpha_T T_g \left(1 - \frac{T}{T_g}\right)], \qquad (4)$$

where $\alpha_T$ is the thermal expansion coefficient, $C$ is a constant, and $\eta_0$ is the extrapolated high temperature limit of the viscosity. As observed, $log_{10}(\eta)$ will scale exponentially with a linearly changing $\lambda$.

The results presented here, obtained by directly fitting the $g(r)$ data, are in contradiction with the conclusions reached by Krausser et al. [2] There may be several reasons for this. In their studies $\lambda$ was determined indirectly from the measured temperature dependence of $G_{inf}$ in the glass, while it is more directly measured here. Further, molecular dynamics (MD) results indicate that the anharmonicity of the atomic potential (correlated with $\alpha_T$) also correlates with the steepness of the repulsive potential,[7] making it difficult to disentangle the two quantities from an analysis of $G_{inf}$ alone. The values for $\lambda$ obtained from $G_{inf}$ by Krausser *et al.* are also one to two orders of magnitude larger than we obtain from fits to $g(r)$. Later studies by the same group[14] found smaller values for $\lambda$ by fitting the $g(r)$ obtained from MD simulations for a Cu-Zr liquid, although it is still an order of magnitude larger than the value determined from experimental scattering data here. This could be due to their use of a modified version of eq. 2,

$$g(r) = g_0(r - \sigma + b)^\lambda \tag{5}$$

where $g_0 b^\lambda \ll 1$. When our data are fit to eq. 5, we obtain systematically larger steepness values ($44 < \lambda < 54$), but the quality of the fit is poorer than from fitting to eq. 2. Importantly, however, if our experimental data are fit to eq. (5) the correlation between steepness and fragility (stronger liquids have larger values of $\lambda$) remains unchanged.

As a further check, $\lambda$ was determined from the $g(r)$ obtained as a function of temperature from a MD simulation of $Cu_{55}Zr_{45}$ (one of the liquids in our experimental study). As shown in fig. 4.a, very good agreement is found between the experimentally measured $g(r)$'s as a function of temperature and those determined from the MD simulation, indicating the accuracy of the simulation. The values of $\lambda$ computed from the experimental data are compared with those from the simulation in fig. 4.b. To within error, they are in very good agreement. Also, like the experimental data the MD data clearly show that $\lambda$ is a function of temperature.

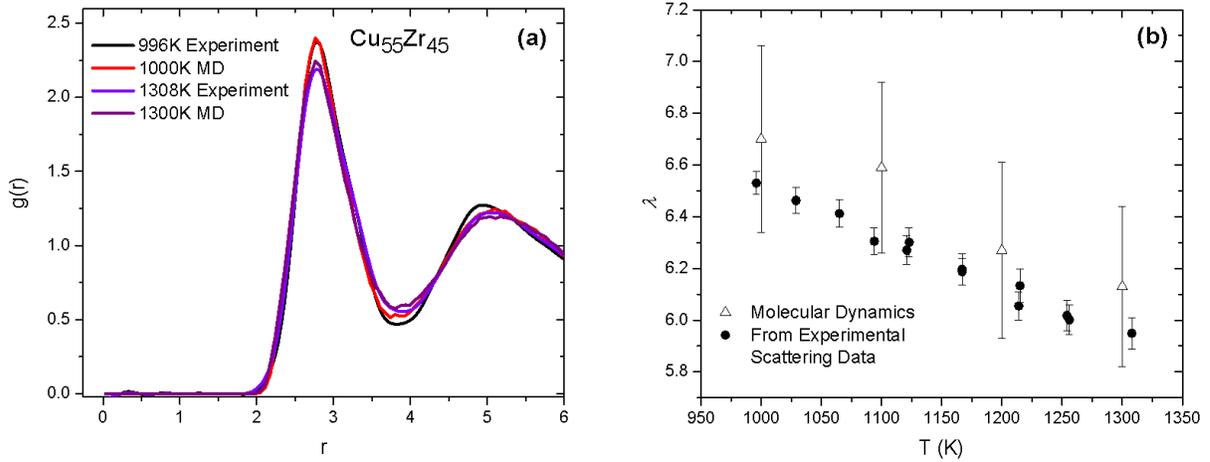

**Figure 4** - (a) A comparison between experimentally measured and MD simulated $g(r)$ data for $Cu_{55}Zr_{45}$, showing excellent agreement (shown at only two selected temperatures for visibility). (b) Values of $\lambda$, obtained by fitting the low-r side of g(r) from scattering data (●) and from MD (△). Both the magnitude and temperature dependences of the two sets of $\lambda$ are in good agreement.

In conclusion, we demonstrate a clear experimental correlation between the nature of the effective atomic potential and the fragility in metallic liquids. In agreement with speculations made almost a half-century ago by Polk and Turnbull,[1] stronger liquids, having a larger viscosity at high temperatures, have a steeper repulsive component. The steepness parameter, $\lambda$, is found to be temperature-dependent, increasing with decreasing temperature, which is consistent with a sharpening of the first peak in $g(r)$. This is also in agreement with MD results of a $Cu_{55}Zr_{45}$ liquid, and with recent MD results for liquids with modified binary Lennard-Jones type potentials,[7-9] and a ternary metallic liquid.[10] The change in $\lambda$ with temperature is larger at high temperature (near $T_A$) for stronger liquids, indicating that these liquids are ordering more rapidly there than are fragile liquids in agreement with recent structural studies of fragility.[23]

It is unclear whether these results are limited to metallic liquids or hold more generally for liquids. An investigation of this could lead to a more general understanding of the microscopic origin of fragility. Also, our results disagree with the conclusions drawn by Mattsson et al. from their studies of colloidal suspensions[4]. Since colloidal suspensions are often taken to be model systems for metallic liquids, these results suggest that the conditions under which this may be true need to be re-examined.

**METHODS**

Alloy liquids were prepared by arc-melting high purity (>99.9%) elements in the appropriate ratio in an argon-atmosphere. Before creating each composition, a TiZr getter was first melted to further reduce oxygen impurities within the arc-melting chamber. All viscosity and X-ray scattering data were obtained using the Washington University Beamline ElectroStatic Levitation (WU-BESL)[24] facility. The liquid viscosity was measured using the oscillating drop technique.[25] A sinusoidal modulating voltage was applied to the levitation field to induce oscillations of the droplet near its resonance frequency. The sinusoidal modulation was then removed and the surface oscillation was allowed to decay. The viscosity was determined from the decay time-constant, $\tau$, which is related to viscosity by $\eta = \frac{\rho R_0}{5\tau}$ where $\rho$ is the density, and $R_0$ the sample radius.

The Arrhenius crossover temperature $T_A$ was obtained by fitting the viscosity to an expression from the avoided critical-point theory (KKZNT)[17]:

$$\log(\eta) = \log(\eta_0) + \frac{1}{T}\left[E_{inf} + T_A B \left[\frac{T_A-T}{T_A}\right]^z \Theta(T_A - T)\right]$$

where $E_{inf}$ is the activation energy at high temperature, $\eta_0$ is the viscosity extrapolated to infinite temperature, $B$ and $z$ are fitting parameters, and $\Theta(T)$ is the Heaviside Function. $E_{inf}$ and $\eta_0$ can be obtained directly from the data and values for $B$ and $z$ have been estimated theoretically.[19]

Structural data were obtained on beam-line 6ID-D at the Advanced Photon Source, Argonne National Laboratory from levitated liquid alloy drops using high energy x-rays (129 keV, 0.0958(6) Å) in a transmission geometry. The scattered intensity was recorded with a GE

Revolution 41-RT area detector over a range of $1 \leq q \leq 15$ Å−1. Background contributions from the Be-window and air scattering were subtracted from the measured intensities. In-house software,[26, 27] designed to correct for absorption, multiple- and incoherent Compton-scattering for samples with a spherical geometry, was used to obtain $S(q)$ as a function of temperature. The pair distribution function $g(r)$ was obtained by a Fourier transform of the $S(q)$ data.

The molecular dynamics simulations were made for Cu-Zr alloys[28] using the embedded atom method (EAM) potential and using LAMMPS.[29] The sample contained 32,000 atoms which were placed in a cubic box with periodic boundary conditions. The NPT(P = 0) ensemble was used, and the pressure and temperature were controlled through a Nose-Hoover barostat and thermostat, respectively. The sample was quenched from an equilibrium liquid at 2000 K to the target temperature at a cooling rate of 1 K/ps. The MD step time was 2 fs.

**ACKNOWLEDGEMENTS**

We thank Doug Robinson for his assistance with the high-energy x-ray diffraction studies at the APS and Anupriya Agrawal, Anup Gangopadhyay, Konrad Samwer and Li Yang for useful discussions. Kelton and Pueblo gratefully acknowledge support by NASA under Grant No. NNX16AB52G and the National Science Foundation under Grant No. DMR 15-06553. Sun gratefully acknowledges support by the Foundation for the Science and Technological Innovation Talent of Harbin (No. 2010RFQXG028). Any opinions, findings and conclusions or recommendations expressed in this article are those of the author(s) and do not necessarily reflect the views of the National Science Foundation or of NASA.

## AUTHOR CONTRIBUTIONS

Kelton conceived of the study. Kelton and Pueblo obtained the experimental data, and Sun carried out the MD simulations. Kelton and Pueblo analyzed all experimental and MD data, and all authors contributed to writing and editing the document.

## COMPETING FINANCIAL INTERESTS

The authors declare no competing financial interests.

**Does the repulsive interatomic potential determine fragility in metallic liquids?**

**Supplementary Information**

**C. E. Pueblo, M. Sun and K. F. Kelton**

Instead of $T_g/T^*$, $T_A/T^*$ was used as a metric for fragility in the high temperature liquids studied here. Of the alloy liquid compositions studied, only four have published values for the fragility parameter, $m$. However, as shown in fig. S1, in those cases, $m$ correlates well with both $T_A/T^*$ and $T_g/T^*$, making either of these quantities a valid measure of fragility in high temperature liquids.

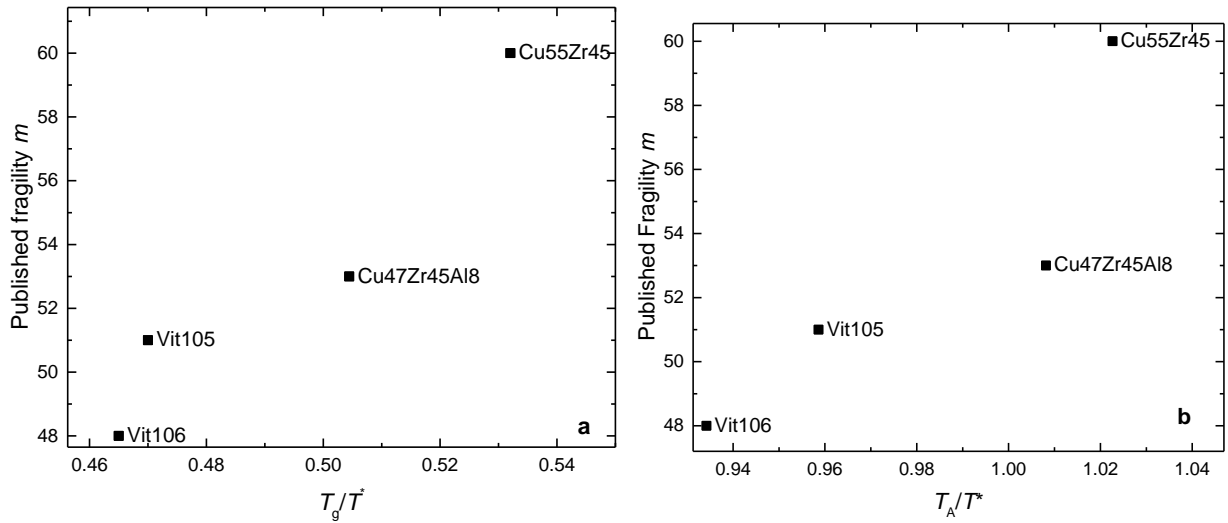

**Supplementary Figure S1** – Literature values of the fragility parameter $m$, as a function of (a) $T_g/T^*$ and (b) $T_A/T^*$.

Since the structural data were obtained from high temperature liquids, the quantity $T_A/T^*$ was used to demonstrate how fragility relates to the steepness of the repulsive part of the potential, obtained by fitting the low-r side of the first peak in g(r). Since not all liquids formed glasses, this is a more useful metric than $T_g/T^*$ to study a wider range of liquids. However, as shown in fig. S1, the same correlation (larger $\lambda$ for stronger liquids) is found in those liquids that form glasses, with measurable values of $T_g$.

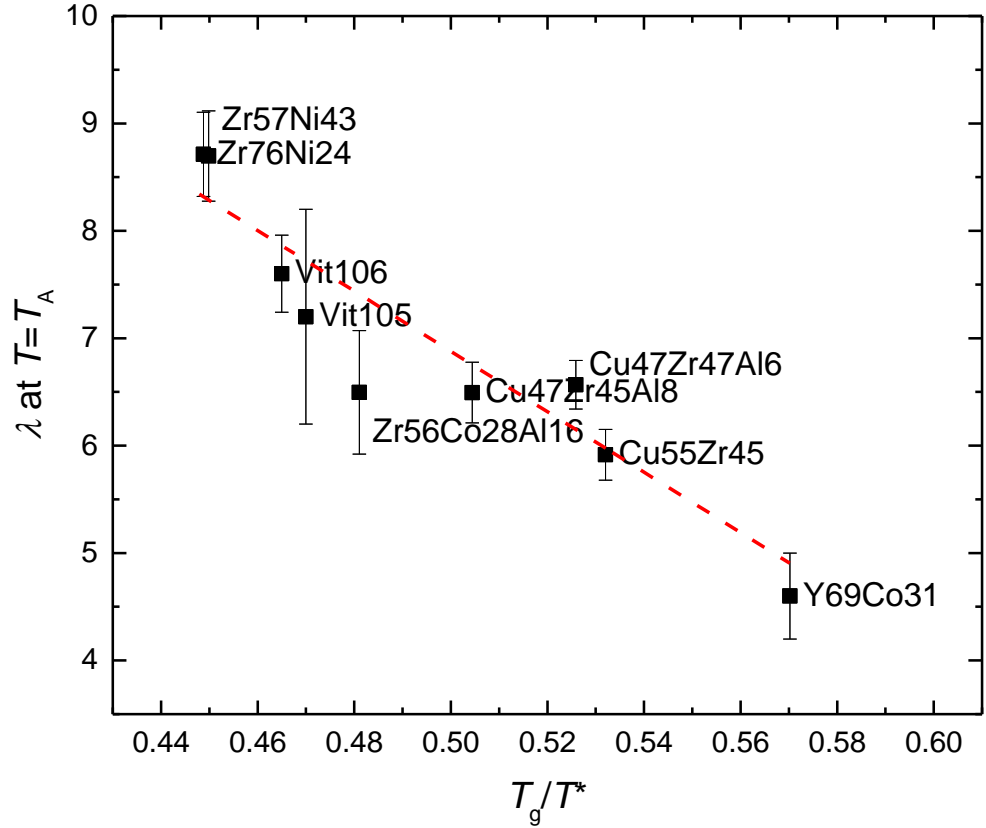

**Supplementary Figure S2** – Steepness parameter, $\lambda$, as a function of $T_g/T^*$, showing the same correlation as when $\lambda$ is expressed as a function of $T_A/T^*$, i.e. stronger liquids have steeper repulsive potentials. $Ti_{38.5}Zr_{38.5}Ni_{23}$ is omitted here (but shown in fig. 3 in the manuscript) since it does not form a glass.

The values used for these comparisons as well as the steepness values, $\lambda$, and their derivatives with temperature are listed in Table S1.

# Table S1

Compiled temperature, and fragility data, with error in parentheses. The values for $T_g$ were obtained from differential scanning calorimeter studies. Most of these were obtained by A. K. Gangopadhyay; those marked with an asterisk were measured by C. E. Pueblo.

| Composition | $T_A$ (K) | $T_g$ (K) | $m$ | $T_A/T^*$ | $\lambda$ (at $T_A$) | $d\lambda/dT$ (K$^{-1}$) |
|---|---|---|---|---|---|---|
| Vit105 | 1329 | 670 | 51[1] | 0.93198 (.00146) | 7.2 (1) | -0.00225 (3.07E-4) |
| Cu55Zr45[a] | 1309 | 681 | 60[2] | 1.02266 (.00764) | 5.915 (.234) | -0.00196 (7E-5) |
| Cu47Zr47Al6 | 1317 | 701 | | 0.988 (.00152) | 6.567 (.227) | -0.0025 (7.61E-5) |
| Vit106 | 1363 | 678 | 48[1] | 0.9342 (.00147) | 7.601 (.359) | -0.00286 (1.47E-4) |
| Zr56Co28Al16 | 1499 | 749 | | 0.96275 (.00133) | 6.496 (.576) | -0.00325 (2.59E-4) |
| Zr57Ni43 | 1231 | 655* | | 0.84547 (.0065) | 8.697 (.421) | -0.00409 (1.76E-4) |
| Cu47Zr45Al8[b] | 1361 | 681* | 53[3] | 1.00815 (.00147) | 6.494 (.282) | -0.00234 (9.02E-5) |
| Y69Co31 | 1208 | 615* | | 1.12 (.03808) | 4.6 (.4) | -0.00115 (1.96E-4) |
| Zr76Ni24 | 1129 | 603* | | 0.84443 (.00443) | 8.712 (.392) | -0.00326 (1.38E-4) |
| Ti38.5Zr38.5Ni23 | 1159 | -- | | 0.88473 (.0069) | 7.658 (.254) | -0.0028 (8.9E-5) |

[a]Fragility $m$ estimated from $Cu_{56}Zr_{44}$

[b]Fragility $m$ estimated from $Cu_{46}Zr_{46}Al_8$

**References**
1. Johnson, W., Na, J. and Demetriou, M., Quantifying the origin of metallic glass formation. *Nat. Commun.* **7**, (2016).
2. Russew, K., Stojanova, L., Yankova, S., Fazakas, E. and Varga, L., presented at the Journal of Physics: Conference Series, 2009 (unpublished).
3. Zhou, C., *et al.*, Structural evolution during fragile-to-strong transition in CuZr (Al) glass-forming liquids. *J. Chem. Phys.* **142**, 064508 (2015).